\documentclass[aip,apl,twocolumn,reprint,amsmath,amssymb,superscriptaddress]{revtex4-1}
\usepackage{graphicx}
\usepackage{dcolumn}
\usepackage{bm}
\usepackage{amsmath}

\begin{document}

\title{Fabrication and characterization of ambipolar devices on an undoped AlGaAs/GaAs heterostructure}

\author{J.C.H. Chen}
\email{jchen@phys.unsw.edu.au} \affiliation{School of Physics,
University of New South Wales, Sydney NSW 2052, Australia}

\author{D.Q. Wang}
\affiliation{School of Physics,
University of New South Wales, Sydney NSW 2052, Australia}

\author{O. Klochan}
\affiliation{School of Physics, University of New South Wales, Sydney NSW 2052, Australia}

\author{A.P. Micolich}
\affiliation{School of Physics,
University of New South Wales, Sydney NSW 2052, Australia}

\author{K. Das Gupta}
\affiliation{Cavendish Laboratory, J J Thomson Avenue, Cambridge CB3 0HE, United Kingdom}
\altaffiliation[Now at ]{Indian Institute of Technology, Bombay, Powai, Mumbai 400 076, India}

\author{F. Sfigakis}
\affiliation{Cavendish Laboratory, J J Thomson Avenue, Cambridge CB3 0HE, United Kingdom}

\author{D.A. Ritchie}
\affiliation{Cavendish Laboratory, J J Thomson Avenue, Cambridge CB3 0HE, United Kingdom}


\author{D. Reuter}
\affiliation{Angewandte Festk\"{o}rperphysik, Ruhr-Universit\"{a}t
Bochum, D-44780 Bochum, Germany}

\author{A.D. Wieck}
\affiliation{Angewandte Festk\"{o}rperphysik, Ruhr-Universit\"{a}t
Bochum, D-44780 Bochum, Germany}

\author{A.R. Hamilton}
\email{Alex.Hamilton@unsw.edu.au} \affiliation{School of Physics, University of New South Wales, Sydney NSW 2052, Australia}


\begin{abstract}
We have fabricated AlGaAs/GaAs heterostructure devices in which the conduction channel can be populated with either electrons or holes simply by changing the polarity of a gate bias. The heterostructures are entirely undoped, and carriers are instead induced electrostatically. We use these devices to perform a direct comparison of the scattering mechanisms of two-dimensional (2D) electrons ($\mu_\textrm{peak}=4\times10^6\textrm{cm}^2/\textrm{Vs}$) and holes ($\mu_\textrm{peak}=0.8\times10^6\textrm{cm}^2/\textrm{Vs}$) in the same conduction channel with nominally identical disorder potentials. We find significant discrepancies between electron and hole scattering, with the hole mobility being considerably lower than expected from simple theory.
\end{abstract}


\maketitle
Modulation-doped $\textrm{Al}_{x}\textrm{Ga}_{1-x}\textrm{As/GaAs}$ heterostructures have formed the starting point for innumerable studies of low dimensional electron and hole systems.~\cite{Beenaker} In modulation doped heterostructures the spatial separation of the dopants and the channels significantly reduces scattering, so that very high electron or hole mobilities can be achieved at low temperatures.~\cite{ModDope} However, the
type of charge carrier is determined by the dopants used during the heterostructure growth, which makes it very difficult to fabricate ambipolar devices that can operate with both electron and hole conduction. The ability to switch seamlessly between electrons and holes in the same device would be of interest for studies of scattering, interaction effects and spin related phenomena, since the two types of charge carriers have very different effective masses, bandstructures and spin properties.

To create ambipolar devices we eschew conventional modulation doping techniques, using instead a gate electrode to populate the channel electrostatically. The challenge with this approach is to make good electrical contact to the 2D electrons or holes in the channel without forming an unwanted contact to the gate electrode which must overlap the ohmic contact. Hirayama \textit{et al}.~\cite{HirayamaGaAs, SiAmbi} used ion-implantation to overcome this problem and fabricate ambipolar devices on AlGaAs/GaAs heterostructures. Thick high Al content AlGaAs diffusion barriers were used to suppress leakage between the top gate and the ohmic contacts. However, a high temperature anneal ($\sim 800^{\circ}$C) is required after the ion implantation to activate the dopants, which is higher than the wafer growth temperature and may have adverse effects on the heterostructure and the carrier mobility.
Here we describe ambipolar devices fabricated without the need for ion implantation and subsequent dopant activation anneals. With this device design we are able to make high quality ambipolar 2D systems, and compare the transport lifetime of electrons and holes formed in the same channel, with the same scattering potential.

\begin{figure}
\includegraphics[width=8.5cm]{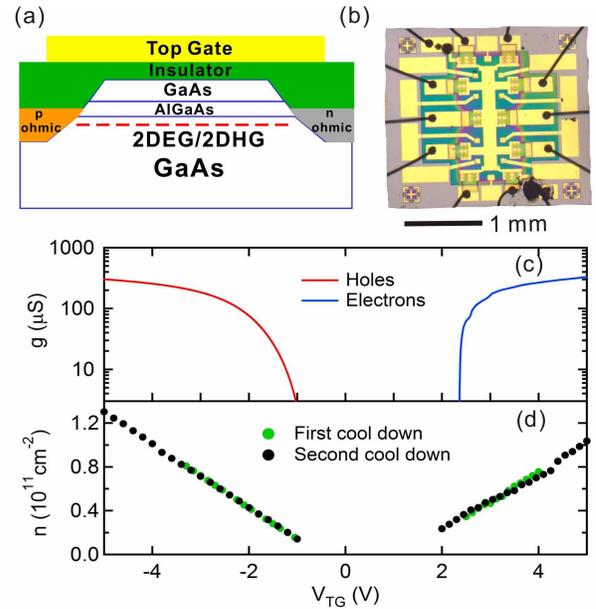}
\caption{(a) Device schematic, showing top gate separated from the ohmics by the polyimide. (b) Optical micrograph of a typical induced device used in this study. (c) Two terminal conductance of electrons ($V_{\textrm{TG}}>0$) and holes ($V_{\textrm{TG}}<0$) through the device at $240$ mK with an excitation voltage of $100 \mu$V. (d) Carrier density as a function of applied top gate bias. The densities are linear in top gate voltage and are highly reproducible between thermal cycles.}
\end{figure}

We have fabricated devices from a number of different undoped heterostructures,  extending the approach described in Refs. \onlinecite{Harrell} and \onlinecite{Sarkozy2D} for making unipolar devices. Ohmic contacts are fabricated by standard optical lithography techniques, so there is no need for ion implantation. Here we present data from wafer B13520 grown by molecular beam epitaxy on a (100) GaAs substrate, although similar data was obtained on other wafers. A $1\mu$m GaAs buffer layer was followed by 300 nm of undoped AlGaAs and capped by 17 nm of GaAs. Standard UV photolithography was used to define the mesa, ohmics, polyimide and top gate patterns. AuBe alloy was used for the hole contacts and NiAuGe for the electron contacts. A 500 nm thick layer of polyimide was used as the insulator to isolate the top gate from the ohmic contacts, while allowing the top gates to overlap the ohmic contacts. The top-gate was deposited on top of the polyimide by thermal evaporation. A schematic of the device is shown in figure 1(a) and a micrograph of a typical ambipolar device is presented in figure 1(b). The  long rectangular gold pattern in the middle of the device is the Ti/Au top gate, which defines the conduction channel.

The device was characterised at $240$ mK using standard lock-in techniques. The two-terminal conductance is shown in Fig. 1(c) as a function of top-gate gate bias $V_{\textrm{TG}}$. The channel is populated with holes for $V_{\textrm{TG}}<0$, and with electrons for $V_{\textrm{TG}}>0$. The threshold voltages for electrons and holes depend on how well the three-dimensional ohmics contact to the 2D charge carriers, which can vary from device to device. In contrast the 2D electron and hole densities were extremely reproducible between thermal cycles on the same device (see Fig. 1(d)), and consistent within $\pm 5\%$ between devices from the same wafer. The range of densities accessible was limited by the requirement to keep the leakage current between gate and ohmic contacts below 1 nA at the high end, and by the threshold voltage of the ohmics at the low end.

Magnetotransport data were taken at different top gate voltages in a perpendicular magnetic field up to 2T. Figure 2 shows typical transport data for both electrons and holes. From the Shubnikov-de Haas oscillations the densities were calculated at each top-gate bias. The densities are plotted in Fig. $1(d)$ and increase linearly with top gate voltage. The slope of $dn/dV_{\textrm{TG}}$ was similar in all devices measured at $2.7 \pm0.1 \times 10^{10}m^{-2}/V$. The slopes were used to calculate the thickness of the polyimide layer for each device, which was consistent with the experimental values measured with a Dektak surface profilometer ($500-600$ nm). The top-gate biases at which the 2D electron and hole densities extrapolated to zero were 1.09~V and -0.52~V respectively, with the 1.61~V difference between these values being very close to the low temperature GaAs band gap of 1.52~eV.

\begin{figure}
\includegraphics[width=8.5cm]{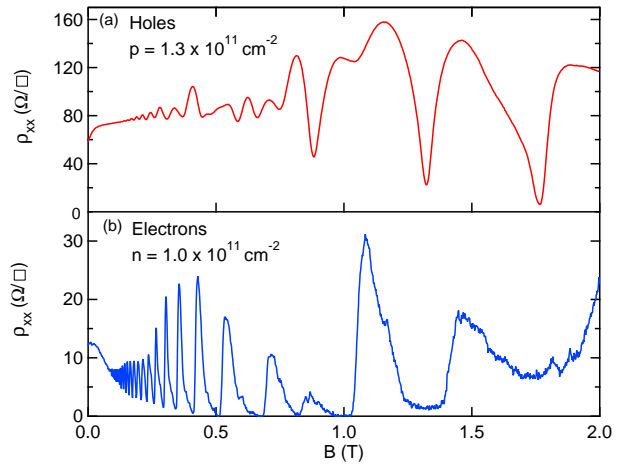}
\caption{Magnetoresistance measured at 240 mK, showing the Shubnikov de-Haas oscillations for similar electron and hole densities.}
\end{figure}

The Shubnikov-de Haas oscillations shown in Fig. 2 are more pronounced for electrons than holes. This is because the hole effective mass is $3-5$ times larger than the electron effective mass. Since the Landau level separation is $\hbar eB/m^{*}$, the effects of disorder broadening and thermal smearing are more severe for holes than electrons at the same measurement temperature.

\begin{figure}
\includegraphics[width=8.5cm]{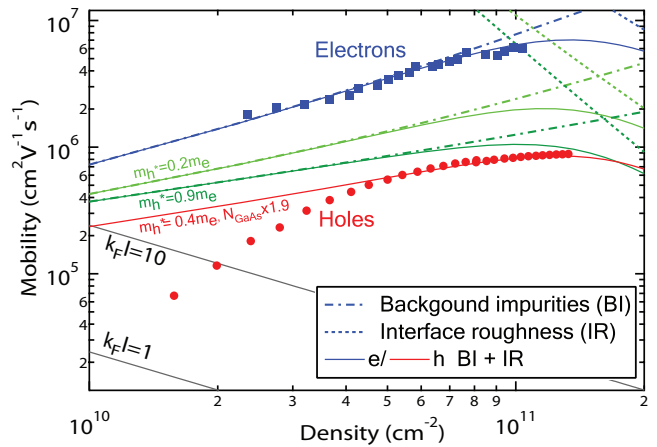}
\caption{Symbols show mobility as a function of density for electrons (squares) and holes (circles). Four sets of calculations of total mobility are shown as solid lines. From the top down the first three are for electrons (blue), holes with $m_{h}^{*} = 0.2m_{e}$ (light green), and holes with $m_{h}^{*} = 0.9m_{e}$ (dark green), all calculated with the same interface roughness and background impurity levels. The fourth solid line (red) is for  holes with $m_{h}^{*} = 0.4m_{e}$ using a higher background impurity level in the GaAs. Dashed and dash-dotted lines are the calculated mobilities accounting only for interface roughness ($\Delta$=2 \AA\, $\Lambda$=4~nm) and background impurities ( $N^{GaAs}_{BI} = 2.42 \times  10^{13} \textrm{cm}^{-3}$, $N^{AlGaAs}_{BI} = 7.25 \times 10^{13} \textrm{cm}^{-3}$). The total mobility includes both contributions. The grey diagonal lines in the lower left corner indicate the region where $1 < k_{F}l < 10$.}
\end{figure}

The carrier density and mobility calculated from the magnetotransport data are plotted in Fig. 3. For both types of charge carriers, the mobility initially increases with carrier density as shown by the squares (electrons) and circles (holes) in Fig. 3. In the conventional picture of transport in high mobility 2D systems this is due to the increase in the Fermi velocity of the charge carriers as the density is increased. At intermediate densities ($10^{11} cm^{-2}$) the mobility saturates, and then stays relatively constant to the highest density measured. This occurs because at higher densities the carriers are pressed against the AlGaAs/GaAs interface, which enhances interface roughness scattering. The peak mobility for electrons is $4\times 10^6 \textrm{cm}^{2}/\textrm{Vs}$ compared to $7.9\times 10^5 \textrm{cm}^{2}/\textrm{Vs}$ for holes at 235 mK.

To compare the scattering mechanisms between electrons and holes, we have modelled the $T=0$ mobility as a function of density with the approach detailed in Refs. ~\onlinecite{Warrick2D, Sarkozy2D, SarahMac, WMak}. The calculations took into account the wavefunction of the charge carriers, screening in the Hubbard approximation, background impurity scattering and interface roughness scattering. Previous modelling of similar heterostructures has shown that the background doping is approximately three times higher in AlGaAs than GaAs,~\cite{WMak, SarahMac} and this ratio was used to model the mobility data for electrons in Fig. 3. The interface roughness is characterized by two parameters, the mean amplitude of the interface roughness $\Delta$ and the roughness correlation length $\Lambda$. The best fit to the measured electron mobility was obtained with $\Delta$=2 \AA\ and $\Lambda$=4~nm, which is comparable to the values used in Ref. ~\onlinecite{Sarkozy2D}. The same values were used to calculate the hole mobility. The grey lines delineate the region $k_{F}l < 10$, where single particle scattering theory begins to break down.

The calculated mobilities are plotted as a function of density in Fig. 3, with blue lines for electrons and red/green for holes. At high densities interface roughness (IR) is the main scattering mechanism, whereas background impurity (BI) scattering dominates at low densities. The total calculated mobilities are shown with solid lines. The best fit for the electron data was obtained with an impurity density of $N^{GaAs}_{BI} = 2.42 \times  10^{13} \textrm{cm}^{-3}$ and $N^{AlGaAs}_{BI} = 7.25 \times 10^{13} \textrm{cm}^{-3}$.

After fitting the electron mobility data, we used the same impurity densities and interface roughness parameters to calculate the expected mobility of the holes, shown as the solid green line in Fig. 3. Since the hole bands are non-parabolic, the hole mass depends both on the heterostructure details and the hole density. We have therefore calculated the expected hole mobility for hole masses of $0.2m_0$ and $0.9m_0$, which are at the extreme limits of values reported in the literature. For all values of the hole mass, the calculated hole mobility (green solid lines in Fig 3) consistently exceeds the measured data. This is an unexpected result, since at first sight the disorder potential should be exactly the same for electrons and holes in the same channel, measured in the same cooldown.

Our calculations show that this discrepancy cannot be due to uncertainties in the hole mass. There are a number of possible origins of the discrepancy. (i) We first look at the mobility at low densities, where the discrepancy between the measured and calculated hole mobility is largest. For $p < 4 \times 10^{10} \textrm{cm}^{-2}$ the measured mobility drops off dramatically, with the slope $d\mu_{h}/dp$ being much larger than the calculations (green lines in Fig. 3). This sharp drop in mobility at low densities has also been observed in other induced hole systems,~\cite{Warrick2D, Lilly} and has been taken as evidence that the system is becoming inhomogeneous with transport crossing over to percolation~\cite{DasSarmaPRL05,ManfraPRL07}. Indeed we obtain a reasonable fit to the 2D percolation expression $\sigma(p)=A(\sigma-\sigma_0)^{4/3}$, with $\sigma_0=1\times 10^{10} cm^{-2}$, for low hole densities (not shown). (ii) At higher densities it is unlikely that percolation and inhomogeneities are relevant, since $k_F l > 100$, yet the measured mobility remains lower than the expected mobility for all hole densities.  It is possible that bandstructure effects, arising from the asymmetric confining potential of the single heterojunction, may play a role. The asymmetric confining potential leads to a Rashba splitting of the heavy hole band into two branches, which appear as if they have different masses~\cite{Eisenstein,Winkler-Habib}. These bandstructure effects were not included in the modelling shown in Fig. 3. However initial calculations using two independent heavy hole bands with $m^{+}_{hh} = 0.9m_{0}$ and $m^{-}_{hh} = 0.2m_{0}$, and neglecting inter-band scattering, did not significantly improve the fit to the measured hole data. This is to be expected since the resulting mobility will lie between the mobilities calculated in the absence of bandstructure effects for $m_h^*=0.2m_0$ and $0.9m_0$ (iii) The third possibility is that the number of background ionised impurities is larger in the GaAs channel when it is occupied with holes than when it is occupied with electrons. This increase in the ionised impurity density could arise when switching from electrons to holes because the Fermi level in the triangular self-consistent quantum well moves from above the conduction band to below the valence band. An approximate doubling of the impurity density in the channel moves the calculated mobility due to background impurities lower, resulting in good agreement with the measured mobility for $p > 5 \times 10^{10} \textrm{cm}^{-2}$ (red line in Fig. 3). However it is difficult to reconcile this increase in backround ionised impurities with the fact that the background doping in the MBE chamber is p-type. Overall, our data suggest that the scattering processes are very different, and considerably more complex, for holes than for electrons.

To summarise we have fabricated ambipolar devices on an undoped AlGaAs/GaAs heterostructure. The devices were robust to thermal cycling, and showed high electron and hole mobilities. The density dependence of the electron mobility was well described by standard Fermi golden rule scattering theory, but the hole data is much more complex, with additional scattering mechanisms involved. Our results show that further work is needed to  understand the scattering properties of 2D holes, even when the disorder potential is well characterised. In the future such devices could be used for studies of the anomalous metallic behaviour in 2D systems, as well as transport in quantum wires and dots.\cite{WarrickNat,OlehAPLWire,OlehHoleDot,SeeQD}

This work was supported by the Australian Research Council under the DP and LX schemes, and by the ARCNN. ARH and APM acknowledge ARC Professorial and Future Fellowships respectively; JCHC acknowledges an Australian Postgraduate Award. We thank U. Z\"{u}licke, R. Winkler and A. Croxall for helpful discussions.

\end{document}